\documentclass[12pt]{iopart}
\expandafter\let\csname equation*\endcsname\relax
\expandafter\let\csname endequation*\endcsname\relax
\usepackage{amsmath}

\usepackage{iopams}
\usepackage[dvips]{graphicx}
\usepackage[dvips]{graphics}
\usepackage{graphicx}
\usepackage{color}
\usepackage{blindtext}
\definecolor{dred}{rgb}{0.75,0,0}
\usepackage{graphicx}
\usepackage{dcolumn}
\usepackage{xcolor}
\usepackage{braket}
\usepackage{physics}
\usepackage{appendix}
\usepackage{hyperref}
\hypersetup{
 colorlinks=true,
citecolor=magenta,filecolor=blue,
 linkcolor=magenta,
 }

\usepackage{graphicx}
\usepackage{dcolumn}
\usepackage{bm}
\usepackage{xcolor}

\definecolor{codegreen}{rgb}{0,0.6,0}
\definecolor{codegray}{rgb}{0.5,0.5,0.5}
\definecolor{codepurple}{rgb}{0.58,0,0.82}
\definecolor{backcolour}{rgb}{0.95,0.95,0.92}

\begin{document}


\title{\textcolor{blue}{Flat bands, edge states and possible topological phases in a branching fractal}} 

\author{Sougata Biswas, Amrita Mukherjee, and Arunava Chakrabarti}
\address{Department of Physics, Presidency University, 86/1 College Street, Kolkata, West Bengal - 700 073, India}

\address{Department of Physics, University of Kalyani, Kalyani,
West Bengal-741 235, India}

\address{Department of Physics, Presidency University, 86/1 College Street, Kolkata, West Bengal - 700 073, India}
\eads{
\mailto{biswassougata3@gmail.com},
\mailto{amritaphy92@gmail.com}, and
\mailto{arunava.physics@presiuniv.ac.in}}

\date{\today}

\begin{abstract}
We address the problem of analytically extracting a countable infinity of flat, non-dispersive bands in a periodic array of cells that comprise hierarchically grown branching Vicsek geometries. The structural  units can, in principle, be of arbitrarily large size. Through a geometric construction, followed by an exact real space renormalization scheme we unravel clusters of compact localized states, corresponding to densely packed groups of flat bands, sometimes in close proximity with the dispersive ones, as the unit cells accommodate Vicsek fractal motifs of higher and higher generations. In such periodic arrays, energy bands close and open up at energies that are calculated exactly, and the precise correlation between the overlap integrals describing the tight binding systems has been worked out. The possibility of a topological phase transition is pointed out through an explicit construction of the edge states, weakly protected against disorder, though it is argued that the typical bulk-boundary correspondence may not hold good as the unit cells grow in size to achieve the true fractal character.
\end{abstract}

\submitto{\JPCM}
\maketitle

\section{Introduction}
\label{intro}
The occurrence of completely flat, non-dispersive energy bands in a wide class of periodically ordered networks, and a variety of topological phase transitions exhibited by a whole lot of low dimensional lattices are being investigated with great interest in present day condensed matter physics.

The flatband networks (FBN) expose a class of localized compact modes, the so called `compact localized states' (CLS), in spite of having a perfect translational order in the growth direction(s). The local geometric arrangements present in such FBN lattice models cause a destructive interference among the excitations travelling in such systems, so that the amplitudes of the excitations (electronic wavefunctions, for example) are non-zero and get trapped in {\it islands} which are separated from each other. The amplitudes vanish completely outside such zones of survival~\cite{mati,sergej1,sergej2}. This is what we usually refer to as the CLS. The CLS' have been proposed as prospective candidates for the storage and transfer of quantum information~\cite{rontgen}, and have been `observed' experimentally in arrays of photonic waveguides, grafted on a substrate using ultrafast laser writing techniques~\cite{rodrigo,travkin,xia1,seba1,seba2,xia2}.

The topological phase transition (TPT) is another exciting area that has been receiving continued attention of the condensed matter physics community for quite some time now. The TPT is different compared to the conventional phase transitions which involve a breaking of symmetry in the `Landau sense'. A TPT is best understood through the canonical Su-Schrieffer-Heeger (SSH) model~\cite{su1,su2}. The SSH model describes a linear dimerized chain in tight binding approximation with a periodically staggered distribution of two overlap integrals, say, $v$ and $w$, alternating periodically. The relative magnitudes of the hopping integrals decide whether the bands will touch or a gap will open up at the Brillouin zone boundary, indicating a metallic system in the first case, and an insulating phase in the latter. A journey from one insulating phase ($v< w$) to the other ($v > w$) has to go through the gap-closing event ($v=w$), and this is the characteristic of a TPT.

The TPT is associated with the existence of a wide variety of symmetries and a topological invariant~\cite{asboth}. In recent times, a wealth of interesting physics came up subsequently through detailed investigation of some non-trivial variants of the SSH model~\cite{dias1,dias2,amrita1,amrita2}, and the state-of the art grafting of photonic lattices have provided the necessary platform to examine the TPT through experiments~\cite{lu,ozawa,seba3,seba4,yang}. 

A generic feature of the topological states of matter is the so called bulk-boundary correspondence (BBC) that speaks of an intrinsic relation between topological bulk and boundary states~\cite{hasan,qi}. This is important, as the BBC basically connects a physically measurable quantity, such as the conductance, to a topological bulk invariant~\cite{asboth}. If there is an insulating bulk, and the topological invariant is non-zero, then cutting the insulator will make the surface conducting.
\begin{figure}[ht]
\includegraphics[width=\columnwidth]{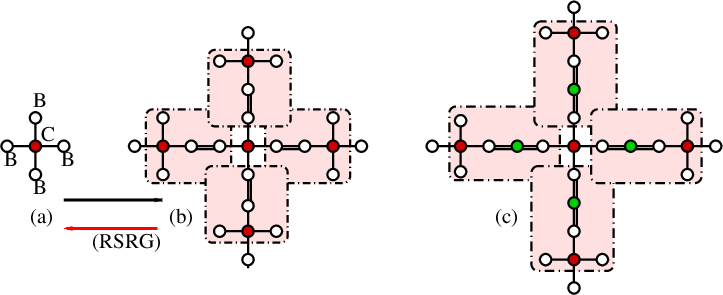}
\caption{(Color online) (a) The first $5$-site cluster, named as the $G_1$ cell in the text, for generating a Vicsek geometry. (b) The second generation ($G_2$) cluster representing a traditional Vicsek fractal (TVF). The inflation is shown by the black arrow, while the red arrow shows the deflation, or the renormalization process. (c) An extended Vicsek geometry in its second generation. The `extra' atomic sites, drawn in green, brings in non-trivial topological changes in the spectrum. In both (b) and (c), the clusters of sites that need to be decimated out for the implementation of the renormalization group ideas (explained in the text) are shown inside the shaded boxes.}
\label{vicsek}
\end{figure}

The above discussion leads us to the precise point where our present work begins. We address ourselves the question of the existence or non-existence of the edge states and their correlation with the topological invariant, for systems where either the distinction between the edge and the bulk is blurred out, by virtue of the lattice construction. or the axis of the inversion symmetry is displaced from the center of the unit cell, causing non-quantized topological invariants. A {\it fractal} space provides one such platform. 

Some fractal lattices, owing to their inherent self-similarity, have been shown to represent a class of FBN's, exposing a countable infinity of non-dispersive energy bands, and the CLS~\cite{biplab1,atanu1,biplab2,atanu2}. In recent times, such geometric fractals are being seriously considered to address the topological issues as well, in both theory~\cite{pai,sarangi}, and experiments, with waveguides grafted on a substrate using ultrafast laser writing techniques~\cite{yang,xie}. These results motivate us to dig deeper into the role of a deterministic fractals substrate in giving birth to the flat bands, and to find out if there is any possibility of any kind of a TPT.

In this communication, we choose a Vicsek fractal (VF) geometry as a prototype of a regular branching fractal network. The Vicsek geometry has been extensively studied in the past in terms of its electronic and other physical properties~\cite{jayanthi1,jayanthi2,hu,goda,bibhas}. There are several interesting aspects of this geometry that trigger our interest in undertaking an in-depth study to explore the existence of CLS' here, and a topological protection of any `edge' eigenmodes that might exist. We can lay down our points of interest, or motivations behind the work as follows:
\vskip .1cm
\noindent

$(i)$ The first point of interest is to work out an analytical prescription to explore the possibility of occurrence of CLS' in a VF structure. This is a non-trivial problem as, knowing the exact energy eigenvalues (at least a subset of them) in a branching fractal structure is difficult, if not impossible, especially when the fractal grows to its thermodynamic limit. This problem has only lately been raised in the literature~\cite{biplab1}. We exploit the inherent self-similarity in a VF geometry (see Fig.~\ref{vicsek}) by applying a real space renormalization group (RSRG) decimation scheme to resolve the occurrence and evolution of the CLS', viz, the distribution of the amplitudes of the localized states among the intricate branches as the VF grows in size. 
\vskip .1cm
\noindent
$(ii)$ As the second and equally important point, we may mention that, the VF geometry is a self-similar stubbed structure (see Fig.~\ref{vicsek}) with open branches. This geometry, with its inherent self-similarity, has not been addressed so far in terms of its topological properties, in contrast to the looped Sierpinski gasket structures that have just started receiving attention~\cite{pai,sarangi,yang,xie}, and providing some unusual characteristics. The branching pattern of a VF, in its thermodynamic limit, blurs out the distinction between the bulk and the boundary, and this naturally motivates us to examine whether the BBC is still observed in such systems, the existence of `edge' states, if any, and their protection against disorder (mimicking a distortion of the substrate). Its worth mentioning that, a branching structure, similar to a VF has only recently been addressed in terms of a one dimensional model of a $2^n$-root topological insulators~\cite{marques}. We believe our results will enrich these primary findings.
\vskip .1cm
\noindent
$(iii)$ Finally, the self-similar VF can be taken as an extension of a stubbed lattice structure that has recently been studied experimentally in terms of the flat band light dynamics~\cite{real}. The VF model, at least up to a few generations, could be achievable in the present day photonic lattice experiments, and the FB's, if they exist, should be observable. The self-similarity, explored through an application of the RSRG methods can easily unravel the FB energy eigenvalues mushrooming out of deeper and deeper scales of length. Light dynamics on such deterministic fractals could be an experimental challenge really, promising exciting physics.

We present results of our investigation based on a traditional Vicsek fractal (TVF) geometry, and an extended Vicsek fractal (EVF) geometry~\cite{dolgushev}. It is found that, both the geometries support CLS at every scale of length. We give exact analytical scheme to extract the CLS eigenvalues. In order to probe the topological issues, we prepare periodic arrays of TVF and EVF unit cells. With an increase in the generation index the cell-size increases. In the limit of an infinitely large cell size,  the results will represent the properties of an infinite fractal.

Here, with a finite sized unit cell placed periodically on a line (the backbone), we find analytically exact conditions for opening or closing of the band-gaps at the Brillouin zone (BZ) boundaries for both a TVF and an EVF geometry. Edge states are found to exist for both of them, though for the TVF they are practically inseparable from the bulk bands, and hence do not show any topological character. However, the edge states for the EVF are clearly observed, and are found to be robust against disorder (at least for moderate disorder). These states are protected by the chiral symmetry. The BBC does not seem to work in such branched structures, as the Zak phase~\cite{zak} for the bands, as the gaps open, is found to be non-quantized.

In section II we discuss the family of flat bands seen in our geometries, and the recursive scheme to work out a plethora of flat band-eigenvalues and their distribution in real space as the cell size grows. Section III analyses  the observations related to the edge states and possibilities of a TPT. In Section IV we draw our conclusions.

\vspace{0.2cm}
\section{The energy bands and the CLS family in a Vicsek geometry}
\subsection{The Flat and the dispersive bands unravelled through an RSRG looking glass} 

\subsubsection{The basic scheme with the minimal sized unit cells}

The first two generations of a traditional Vicsek fractal  and the second generation extended Vicsek fractal are shown in Fig.~\ref{vicsek} (a), (b), and (c) respectively. 

We work using the standard form of the tight binding Hamiltonian, 
\begin{equation}
    H = \sum_j \epsilon_j c_j^\dag c_j + \sum_{j,k} c_j^\dag c_k + h.c.
    \label{ham}
\end{equation}
The `on-site' potential $\epsilon_j$ is chosen as $\epsilon_c$ for the sites at the cluster-center (red colored), and as $\epsilon_b$ for all other sites (empty/white in Fig.~\ref{vicsek}(a)). The `extra' site occurring in the EVF is shown as a green colored site in Fig.~\ref{vicsek}(c)). The on-site potential at this green colored site will be symbolized as $\epsilon_a$. The intra-cluster hopping integrals is taken everywhere as $t$, and the clusters are connected by a hopping $T$, shown by a double bond in (a), (b) and (c) in Fig.~\ref{vicsek}. 

\begin{figure}[ht]
\includegraphics[width=0.8\columnwidth]{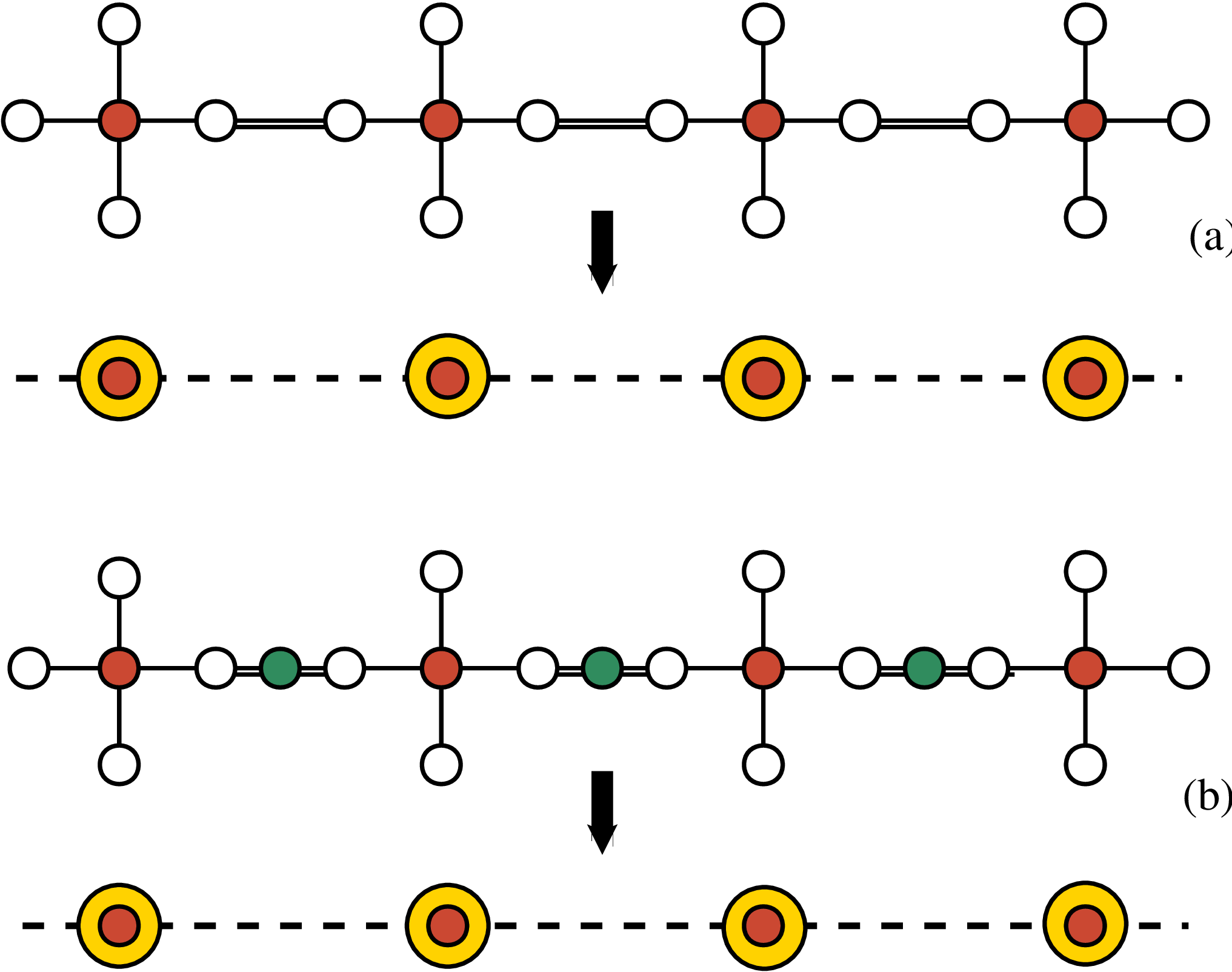}
\caption
{(Color online) Decimation of an array of a TVF (a)  and an EVF (b)  into linear periodic chains. The clusters of sites marked by empty circles have been decimated, and the red sites, encircled by golden color are retained after the decimation.} 
\label{renorm}
\end{figure}

Since our aim is to analyse the possible existence of the FB's and a TPT, we examine the spectral properties of periodic arrays of `unit cells' that are composed of Vicsek cluster-geometries. We will explicitly present results for the minimal sized clusters, and the prescription given by us will allow for extraction of all relevant results when the unit cell grows arbitrarily large, eventually resembling an true VF geometry. 

The implementation of the RSRG is best understood with the simplest possible array, made out of the first generation ($G_1$) clusters of either a TVF or an EVF. 
Fig.~\ref{renorm} shows such an array made out of the first generation ($G_1$) TVF for a `traditional' Vicsek fractal (TVF), and an `extended' Vicsek fractal (EVF)~\cite{dolgushev} in parts (a) and (b) respectively. The central idea is easily understood using these basic units. Extension to higher generation ($G_n$) arrays is straightforward.

We use the discrete version of the Schr\"{o}dinger equation, viz, the `difference equations' that are a set of linear algebraic equations, 
\begin{equation} 
(E - \epsilon_j) \psi_j = \sum_k t_{jk} \psi_k
\label{diff}
\end{equation}
where, $\psi_j$ is the amplitude of the wavefunction on the $j$-th site, and the summation runs over the nearest neighbors of $j$. The nearest neighbor hopping integrals are
$t_{jk}=t$ or $T$ for intracluster, and intercluster connections respectively. 

Using Eq.~\eqref{diff} we decimate, in Fig.~\ref{renorm} (a) and (b), all the atoms {\it except} the red colored central sites and cast the $G_1$ arrays of both the TVF and the EVF into purely linear chains of atoms, now represented by the solid red circles dressed in yellow. These are our `final' 1-d chains that we will be eventually using to extract the FB's and the dispersive bands. 

The decimation, which simply means eliminating the amplitudes $\psi_j$ for the selected set of sites (the empty circles in Fig.~\ref{renorm}(a), and the empty and the green circles in Fig.~\ref{renorm} (b)) and retaining the amplitudes on the red colored sites only,  results in the {\it renormalized} values of the on-site potentials and the nearest neighbor hopping integrals on the chains (made up of the red dots embedded in yellow). They are given by,
\begin{eqnarray}
\tilde{\epsilon}_{TVF} & = & \epsilon_c + \frac{2t^2}{E-\epsilon_b} +
 \frac{2 (E-\epsilon_b) t^2}{[(E-\epsilon_b)^2 - T^2]} \nonumber \\
\tilde{t}_{TVF} & = & \frac{T t^2 (E-\epsilon_b)}{(E-\epsilon_b) [(E-\epsilon_b)^2 - T^2]}
\label{chaintvf}
\end{eqnarray}
for a $G_1$-cell TVF-array, and 
\begin{eqnarray}
\tilde{\epsilon}_{EVF} & = & \epsilon_c + 
 \frac{2 t^2}{E-\epsilon_b} \left [2 + \frac{T^2 (E-\epsilon_b)}{\Delta (E-\epsilon_b)} \right ] \nonumber \\
\tilde{t}_{EVF} & = & \frac{T^2 t^2 (E-\epsilon_b)}{(E-\epsilon_b)^2 \Delta} 
\label{chainevf}
\end{eqnarray}
for a $G_1$ cell EVF array respectively. In the last equation,  $\Delta = (E-\epsilon_a)(E-\epsilon_b) - 2 T^2$. It should be appreciated that cancellation of the common factor $(E-\epsilon_b)$ is  avoided, as cancelling the factor eliminates the possibility of having an eigenvalue $E=\epsilon_b$ in the spectrum.

It is now simple to work out that the dispersion relations for the effective 1-d chains shown in Fig.~\ref{renorm}(a) and (b). The dispersion relation for a TVF array and an EVF array are written as, 
\begin{eqnarray}
   E = \tilde{\epsilon}_{TVF} + 2 \tilde{t}_{TVF} \cos~ka' \nonumber \\
   E = \tilde{\epsilon}_{EVF} + 2 \tilde{t}_{EVF} \cos~ka'
   \label{disp}
  \end{eqnarray}
 Here, `$a'$' symbolizes the appropriate effective lattice spacings of the renormalized 1-d chains (taken as unity later, without losing any physics) and $k$ represents the wave vector. 
  
  Using Eq.~\eqref{chaintvf}, Eq.~\eqref{chainevf}, and Eq.~\eqref{disp}, the dispersion relations for both the TVF and the EVF chains, now comprising only of the yellow-encircled red colored atoms, are recast as,
\begin{equation}
(E - \epsilon_b)~ \xi_{TVF}  =  0 
\label{distvf}
\end{equation}
and, 
\begin{equation}
(E - \epsilon_b)~ \xi_{EVF}  =  0 
\label{disevf}
\end{equation}
for the TVF and the EVF chains respectively. Here, 
\begin{eqnarray}
\xi_{TVF} & = & (E-\tilde{\epsilon}_{TVF}) [(E-\epsilon_b)^2-T^2] - 2 T t^2 \cos~ka' \nonumber \\
\xi_{EVF} & = &  (E - \epsilon_b) (E-\tilde{\epsilon}_{EVF}) \Delta - 2 t^2 T^2 \cos~ka' 
\label{bands}
\end{eqnarray}
\begin{figure}[ht]
\centering
(a)\includegraphics[width=0.4\columnwidth]{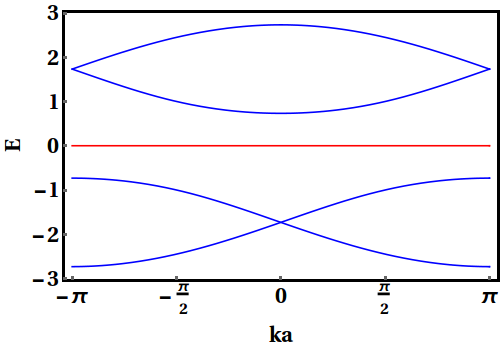}
(b)\includegraphics[width=0.4\columnwidth]{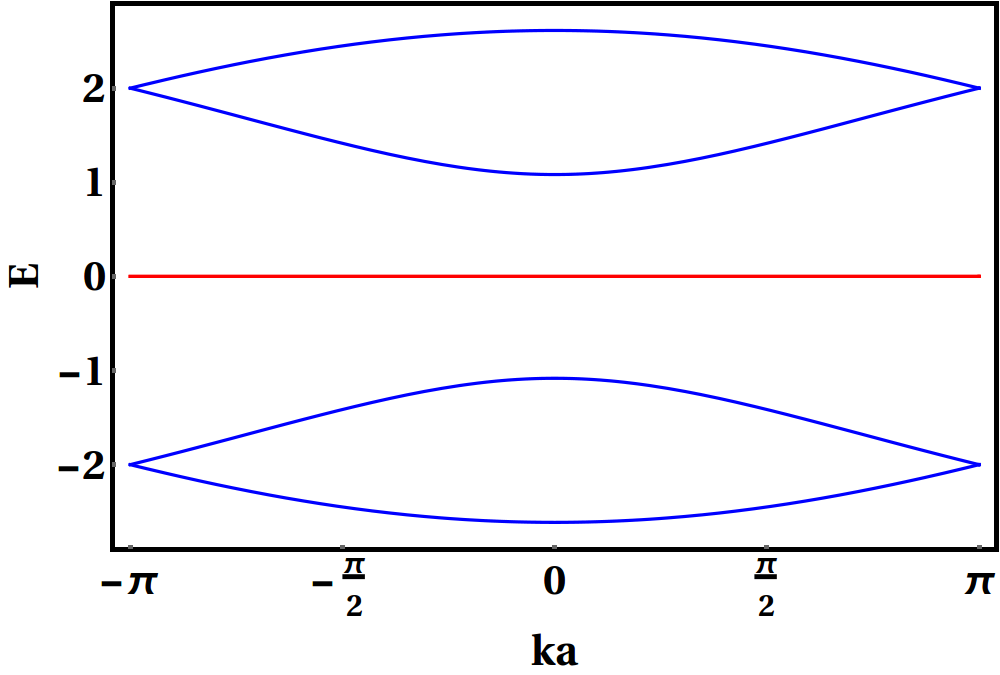}\\
(c)~~\includegraphics[width=0.4\columnwidth]{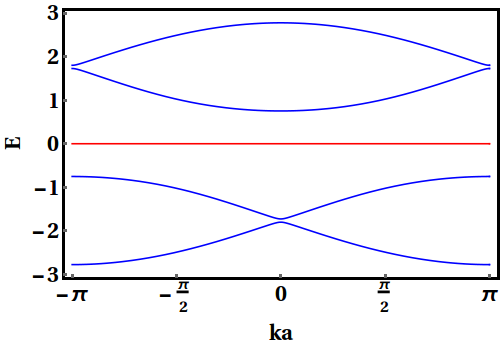}
(d)~~\includegraphics[width=0.4\columnwidth]{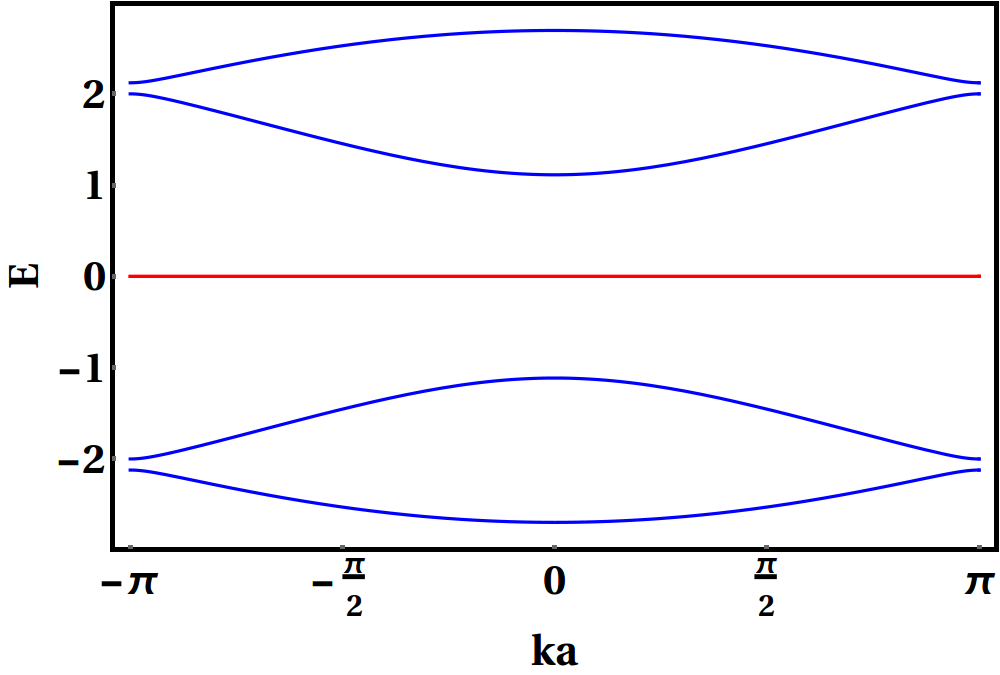}
\caption{(Color online) Energy-wave vector (E vs. ka) dispersion relation for a traditional Vicsek array (a, c) and an extended Vicsek array (b, d), both with the $G_1$ cluster as a unit cell. We have chosen (a) $t = 1$, $\epsilon_b =\epsilon_c=0$, $T=\sqrt{3}$ and (c) $t = 1$, $\epsilon_b =\epsilon_c=0$, $T=1.8$, (b) $t = 1$, $\epsilon_b =\epsilon_c=\epsilon_a=0$, $T=\sqrt{2}$, and (d) $t = 1$, $\epsilon_b =\epsilon_c=\epsilon_a=0$, $T=1.5$. For (a, c) the flat band (red color) is non-degenerate while it is doubly degenerate (red color) for (b, d).}  
\label{disp1}
\end{figure}
\begin{figure}[ht]
\centering
(a)\includegraphics[width=0.44\columnwidth]{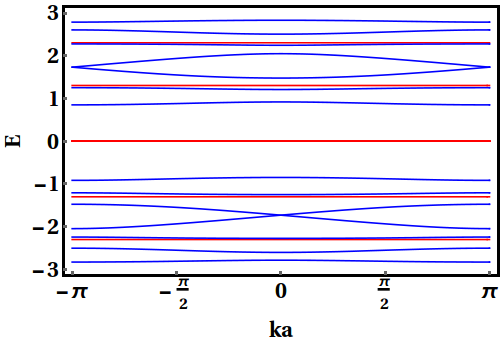}          
(b)\includegraphics[width=0.44\columnwidth]{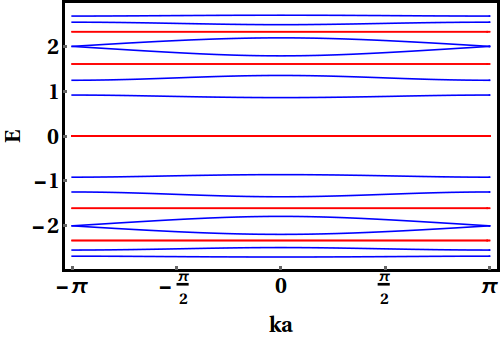}\\
(c)~~\includegraphics[width=0.44\columnwidth]{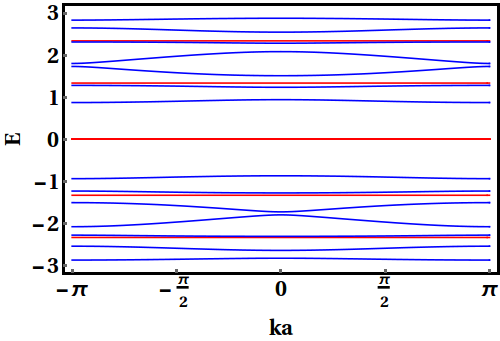}
(d)~~\includegraphics[width=0.44\columnwidth]{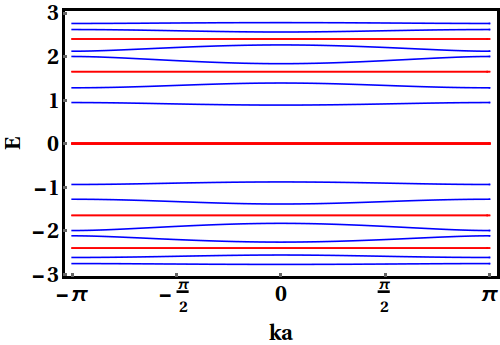}
\caption{(Color online) Energy-wave vector (E vs. ka) dispersion relation for a traditional Vicsek array (a, c) and an extended Vicsek array (b, d), both with the $G_2$ cluster as a unit cell. We have chosen (a) $t = 1$, $\epsilon_b =\epsilon_c=0$, $T=\sqrt{3}$ and (c) $t = 1$, $\epsilon_b =\epsilon_c=0$, $T=1.8$ (b) $t = 1$, $\epsilon_b =\epsilon_c=\epsilon_a=0$, $T=\sqrt{2}$  and (d) $t = 1$, $\epsilon_b =\epsilon_c=\epsilon_a=0$, $T=1.5$. For both cases flat bands are highlighted by red color. }  
\label{disp2}
\end{figure}

From Eq.~\eqref{distvf} and Eq.~\eqref{disevf} one can easily identify the FB occurring at $E=\epsilon_b$ for both a TVF array and an EVF array, while the dispersive bands for both the systems can be extracted from the set of Eqs.~\eqref{bands}.

\subsubsection{Dealing with larger and larger unit cells}

This decimation scheme immediately suggests that, such an array of VF clusters at any $n+1$-th generation, acting as the unit cell of the linear array, will give rise to multiple FB's, existing at the energy eigenvalues that are the solutions of the equation $E-\epsilon_{b}(n) = 0$. Here $n$ stands for the number of decimation steps. This is due to the fact that any basic $5$-site plaquette of a TVF or an EVF cell can be thought of as an $n$ times renormalized version of an $n+1$-th generation VF cluster. If we began with an $n+1$-generation cluster, then after renormalizing it $n$ times, we could arrive exactly into an effective $5$-site unit cell, that formed the array shown in Fig.~\ref{renorm}. In this renormalized $5$-site cluster the effective on-site potentials and the hopping integrals will be functions of energy $E$ of the incoming excitation, the degree of complexity of the functional dependence of course, depending on the stage of renormalization $n$.

The family of dispersive bands will be found as solutions of the equations $\xi_{TVF} (n)=0$, and $\xi_{EVF}(n)=0$ for the linear arrays comprising the $G_{n+1}$ TVF and EVF unit cells. To RSRG scheme depicting a second generation TVF renormalizing to its earlier version, the first generation TVF is presented in Fig.~\ref{vicsek} (a). The green {\it forward} arrow shows the inflation of a $G_1$ cell into a $G_2$ cell. The red arrow, drawn {\it backward} explains the RSRG transformation from $G_2$ to $G_1$. A similar decimation scheme can easily be implemented for the EVF cluster-array.

The parameters $\epsilon_b(n)$, $\epsilon_c(n)$, $t(n)$ and $T(n)$, and hence, both $\xi_{TVF(EVF)} (n)$ are easily obtained from the RSRG recursion relations. In literature~\cite{bibhas} such recursion relations are already available for a TVF. Still, for the sake of completeness, we provide them below.

\begin{figure}[ht]
\includegraphics[width=\columnwidth]{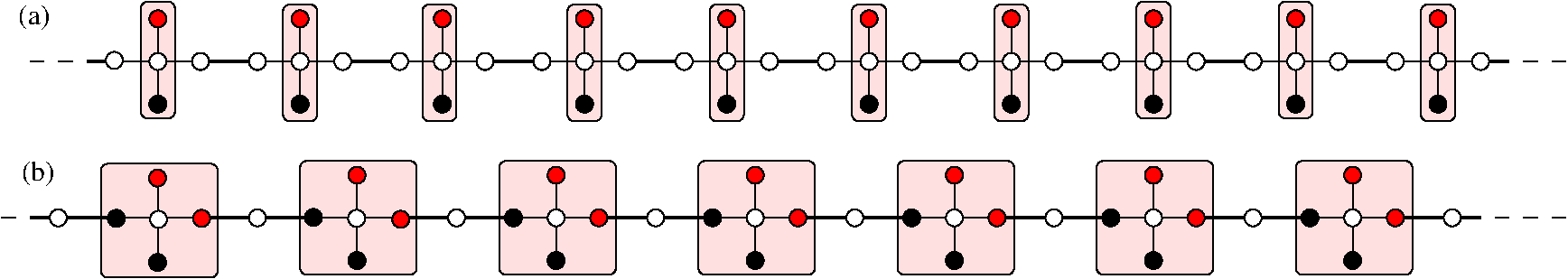}
\caption
{(Color online) Amplitude distribution of a CLS wave function at an energy $E = \epsilon_b $ for (a) traditional  and (b) extended Vicsek array with $ G_1 $ unit cell respectively. In both the figures the zero-amplitudes are marked by white color sites. While red and black color sites indicate non-zero-amplitudes +1, -1 respectively. The compact localized states (CLS) are highlighted by pink color box. The CLS for the extended Vicsek geometry is doubly degenerate, and we have shown just one here.}
\label{cls1}
\end{figure}

\begin{eqnarray}
\epsilon_{b,n+1} & = & \epsilon_{b,n} + \frac{t_{n}^2 [E-\epsilon_{b,n}] \alpha_{n}}
{\alpha_{n} \beta_{n} - t_{n}^2 (E-\epsilon_{b,n})^2}
\nonumber \\
\epsilon_{c,n+1} & = & \epsilon_{c,n}+\frac{4 t_{n}^2 (E-\epsilon_{b,n}) \gamma_{n}} {(E-\epsilon_{b,n})^2 \gamma_{n}^2 - T_{n}^2 [(E-\epsilon_{b,n}) (E-\epsilon_{c,n}) - 2 t_{n}^2]} \nonumber \\
t_{n+1} & = & \frac{t_{n}^3 T_{n} (E-\epsilon_{b,n})}
{\alpha_{n} \beta_{n} - t_{n}^2 (E-\epsilon_{b,n})^2} 
\nonumber \\
T_{n+1} & = & T_{n}
\label{recur}
\end{eqnarray}
where, $\alpha_{n} = (E-\epsilon_{b,n})^2 - T_{n}^2$, 
$\beta_n=(E-\epsilon_{b,n}) (E-\epsilon_{c,n}) - 2 t_n^2$, and $\gamma_n =(E-\epsilon_{b,n}) (E-\epsilon_{c,n}) - 3 t_n^2 $. $n=0$ refers to the bare, `un-renormalized' VF cluster.
The inter-cluster overlap integral $T$ does not change under the RSRG operation.  For an EVF, the RSRG recursion relations have been obtained using a similar rescaling of the lattice, but now the algebra becomes a little more involved, and we skip writing them explicitly here, and provide the details in the appendix.


\subsection{Results from a direct diagonalization of the Hamiltonians}

Let us now check, if the above argument really works. We write down the kernels of the Hamiltonian for the unit cell comprising a $G_1$-TVF (Fig.~\ref{vicsek}(a)) for example, and of a $G_1$-EVF cell subsequently. The kernels read,
 
\begin{equation}
\hat{\mathcal{H}}_{TVF}(k) = \left[ \begin{array}{cccccccccccccccc}
\epsilon_b & t & 0 & 0 & T e^{-ika} \\
t & \epsilon_c & t & t & t \\
0 & t & \epsilon_b & 0 & 0 \\
0 & t & 0 & \epsilon_b & 0 \\
T e^{ika} & t & 0 & 0 & \epsilon_b
\end{array}
\right ] 
\label{ham-normal}
\end{equation}
and, 
\begin{equation}
\hat{\mathcal{H}}_{EVF}(k) = \left[ \begin{array}{cccccccccccccccc}
\epsilon_b & t & 0 & 0 & 0 & T e^{-ika} \\
t & \epsilon_c & t & t & t & 0\\
0 & t & \epsilon_b & 0 & 0 & 0\\
0 & t & 0 & \epsilon_b & 0 & 0\\
0 & t & 0 & 0 & \epsilon_b & T \\
T e^{ika} & 0 & 0 & 0 & T & \epsilon_a
\end{array}
\right ] 
\label{ham-extended}
\end{equation}

Diagonalization of the above Hamiltonians yields the eigenvalues as functions of the wave vector $k$, and the band diagrams (dispersion relations) are presented in Fig.~\ref{disp1}. Panels $(a)$ and $(c)$ in the figure displays the $E-k$ diagram for a linear periodic array comprising the $G_1$ cell of a TVF. For a choice of $\epsilon_b=\epsilon_c=0$ and $t=1$, the bands  close at $E=\sqrt{3}$ if we choose $T=\sqrt{3}$ (both in units of $t$). Detuning $T$ from $\sqrt{3}$ to $\sqrt{3}\pm \delta$, a gap opens at the Brillouin zone (BZ) boundary around $E=\sqrt{3}$. We can thus go from one insulating phase to another through a gap closure, and the question of a TPT becomes pertinent here. At $k=0$, a Dirac cone appears.

Figure panels $(b)$ and $(d)$ show similar results for the array of the $G_1$ EVF cells. The choice of $T=\sqrt{2}$ is shown to close the gaps at the BZ boundaries at $E = \pm 2$, and the gaps open up around the same energies as we deviate from the special choice of $T$. The explanation for such  special values of $T$ will be given shortly. 
Fig.~\ref{disp2} shows the intricacy of the bands and existence of multiple flat bands as we analyse periodic arrays of second generation ($G_2$) clusters of both TVF and EVF. All the bands, both dispersive and the flat ones, obtained from a direct diagonalization of the Hamiltonians, match exactly with the results predicted by the RSRG decimation scheme mentioned above.

The flat bands occur at $E=0$ in all the fours panels in Fig.~\ref{disp1} and Fig.~\ref{disp2}. The energy is actually $E=\epsilon_b$, and $\epsilon_b=0$ in all the calculations here. The choice of $E=\epsilon_{b,n}$ at any $n$-th stage of renormalization leads to a fixed point of the parameters space, as given by Eq.~\eqref{recur} and Eq.~\eqref{recurevf} for $n+1$-th stage onwards, for both TVF and EVF. The FB's arising in arrays with any $n$-th generation TVF or EVF are then naturally retained in the higher order spectra.

This is precisely what we expect from our exact RSRG analysis as presented in the last subsection. The eigenstates corresponding to such flat bands are distributed over clusters of finite extent, that are separated from each other, forming {\it islands} of amplitudes, by sites where the amplitude is zero. We display the distribution on the TVF and EVF $G_1$ arrays, which are the simplest of course, in Fig.~\ref{cls1}. As the size of the unit cells grow, using $G_2$ or $G_3$ or any higher order clusters, the CLS amplitudes spread out in the branches, and the CLS clusters grow in size as well, but still, at any scale of length one cluster remains isolated from the neighboring clusters. 

\begin{figure}[ht]
\includegraphics[width=0.8\columnwidth]{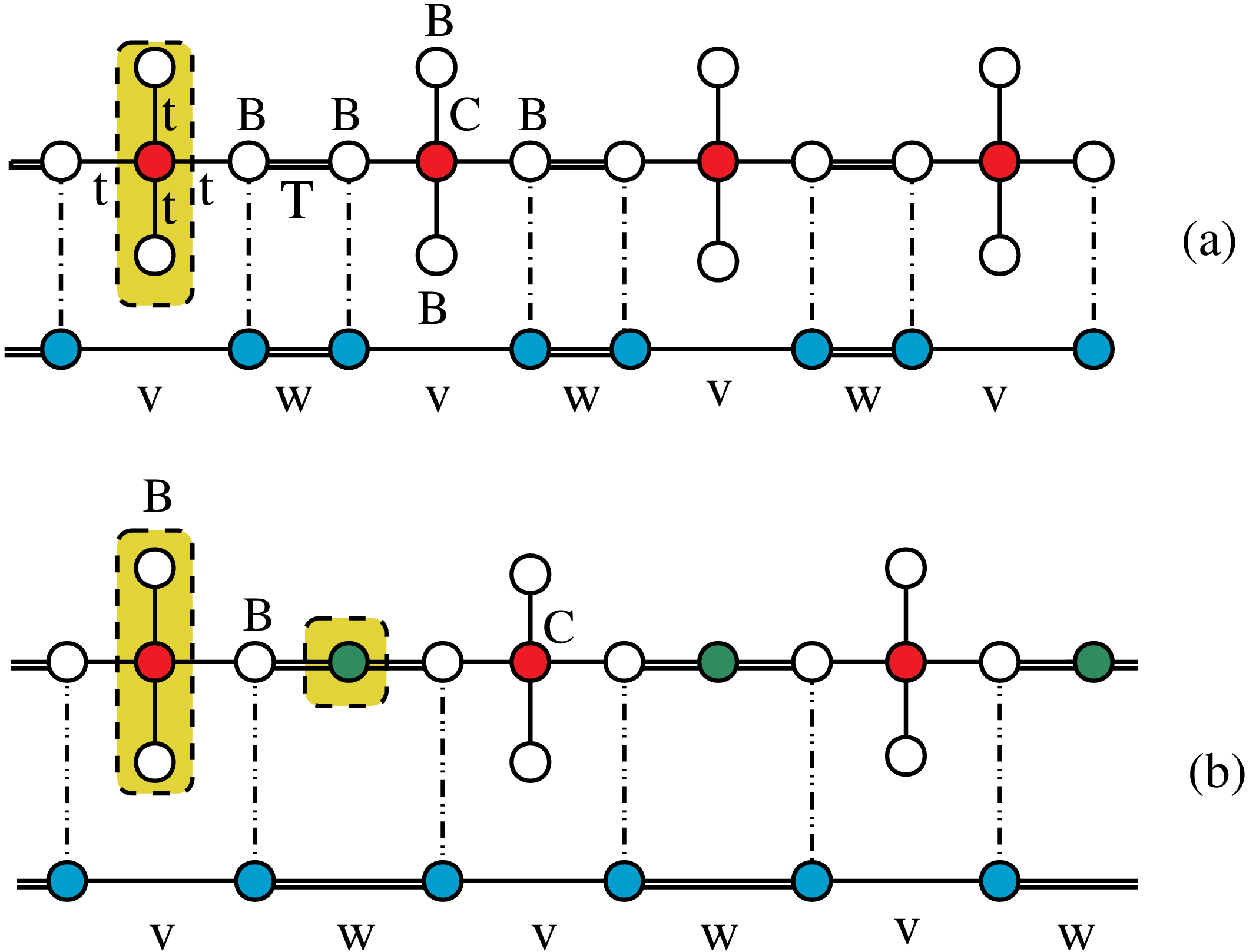}
\caption
{ (Color online) The sequence of decimation steps while going from an array of (a) $G_1$ cells of a TVF and (b) $G_1$ cells of an EVF  to an effective SSH chain (array of cyan colored sites)  with energy-dependent staggered hopping(s). The green sites are the {\it extra} sites in the EVF geometry. In each figure the sites or the blocks of sites decimated along the chain are exemplified by the shaded boxes. The hopping integrals in the effective SSH chain are marked as $v$ and $w$, and the explicit expressions for all of them are given in the text.} 
\label{RG2SSH}
\end{figure}

\vskip  0.2in
\subsection{Closing and opening of the bands : an exact determination of the criteria}
\vspace{0.2cm}

We now discuss how to obtain a specific correlation between the hopping integrals $t$ and $T$ that can close the band-gap at the BZ boundary.

For this we use the difference equation version of the Schr\"{o}dinger equation, viz, and an RSRG decimation scheme again, in a different way.

In Fig.~\ref{RG2SSH}(a) we demonstrate the procedure by considering the simplest possible array, a string of $G_1$ TVF clusters, coupled periodically. In a sequence of decimation steps we eventually `eliminate' the sites appearing within the shaded box (shown only in one or two places), namely, the central red colored site and the white (empty) ones in the stub, all along the chain.
This decimation maps an array of VF clusters, of any generation, into an effective SSH chain, comprising a staggered distribution of a pair of energy-dependent hopping integrals $v$ and $w$, and a uniform {\it effective} energy dependent on-site potential, written as $\epsilon_\alpha$. 
The resulting {\it effective} SSH chain is shown as an array of blue colored sites ($\epsilon_\alpha$). In (b) the same principle is applied to an array of an EVF-cluster in $G_1$, where the eliminated (decimated atoms/groups of atoms) are again shown in shaded boxes. 

We discuss below the central idea by referring to the TVF cluster-array. The discussion of the EVF array follows an exactly similar trail, but obviously with more involved mathematical expressions. 

The renormalized parameters for the effective SSH chain in the first case of a TVF-array are explicitly given by,
\begin{eqnarray}
\epsilon_\alpha & = & \epsilon_b + \frac{t^2 (E-\epsilon_b)}{(E-\epsilon_b) (E-\epsilon_c) - 2 t^2} \nonumber \\
v & = &  \frac{t^2 (E-\epsilon_b)}{(E-\epsilon_b) (E-\epsilon_c) - 2 t^2} \nonumber \\
w & = & T 
\label{effectivessh}
\end{eqnarray}
Now, remembering the SSH result, it is easy see that, the bands will close at the Brillouin zone boundary for $E=\epsilon_\alpha$, provided we set $v=w$. 
A simple algebra will show that, this results in an energy eigenvalue $E=\epsilon_b+T$, {\it provided} we set 
\begin{equation}
    T = \frac{(\epsilon_c-\epsilon_b) \pm \sqrt{(\epsilon_c-\epsilon_b)^2 + 12 t^2}}{2}
    \label{Tvalue}
\end{equation}

\begin{figure}[ht]
\centering
(a)\includegraphics[width=0.44\columnwidth]{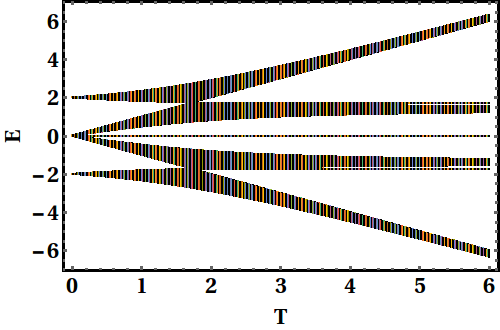}
(b)\includegraphics[width=0.44\columnwidth]{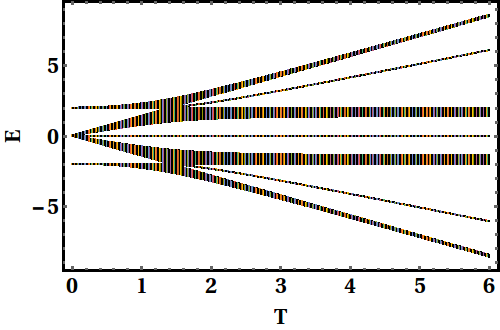}\\
(c)~~\includegraphics[width=0.44\columnwidth]{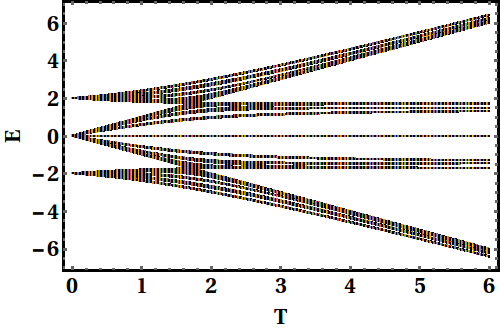}
(d)~~\includegraphics[width=0.44\columnwidth]{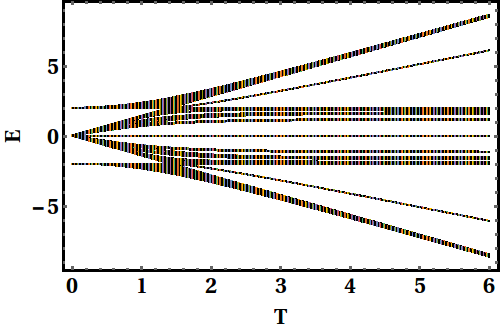}
\caption{(Color online) 
Energy spectra for an array of  TVF (a,c) and  EVF (b,d) clusters with $G_1$ and $G_2$ structures as unit cells, as the inter-cluster coupling $T$ varies. Emergence of the gap states in the EVF are apparent, while the TVF array does not really show any prominent gap state decoupled from the bulk bands.
We have used open boundary conditions for $N_x=40$, where $N_x$ denote the number of unit cells taken along the $x$-direction. The values of the parameters are chosen as $t = 1$, $\epsilon_b =\epsilon_c = 0$.}
\label{emergence}
\end{figure}

For example, if we work with a periodic array of the $G_1$ clusters of a TVF, then a choice of $\epsilon_b=\epsilon_c=0$ and $t=1$ will result in a closing of the band gap at $E=\pm \sqrt{3}$ if we set $T=\pm \sqrt{3}$. 

A similar decimation method, used in case of a $G_1$ array of an EVF unit cell finds the bands closing at {\it two} energy eigenvalues now, at $E=\pm 2t$, when we set $T=\sqrt{2}t$. This condition is easily obtained by noting that, the effective SSH chain, derived out of decimation of a $G_1$ array of an EVF has the parameters, 

\begin{eqnarray}
\epsilon_\alpha & = & \epsilon_b +  \frac{t^2 (E-\epsilon_b)}{(E-\epsilon_b) (E-\epsilon_c) - 2 t^2} + \frac{T^2}{E-\epsilon_a} \nonumber \\
v & = & \frac{t^2 (E-\epsilon_b)}{(E-\epsilon_b) (E-\epsilon_c) - 2 t^2} \nonumber \\
w & = & \frac{T^2}{E-\epsilon_a}
\label{evfparameters}
\end{eqnarray}

where, $\epsilon_\alpha$ now stands for the uniform, energy-dependent on-site potential of the effective SSH chain derived out of the $G_1$ array in the EVF category. This is exactly what we see in Fig.~\ref{disp1}, in all the four panels.

At this point it is worth mentioning that, the self-similar growth pattern of a TVF or an EVF naturally opens up the possibility of extracting, in principle, a countable infinity of `band-closing' energies. The RSRG decimation becomes a useful tool in this respect. To clarify, if we imagine that the five-site TVF clusters in Fig.~\ref{RG2SSH}(a) are actually obtained by decimating out a subset of sites in a $G_{n+1}$ TVF, as shown in Fig.~\ref{vicsek} (a) for $n=1$, then the `gap-closing' energies will be the common roots of two equations, namely,
\begin{eqnarray}
(E-\epsilon_{b,n}) (E-\epsilon_{c,n}) - 3 t_n^2 & = & 0 \nonumber \\
E - \epsilon_{b,n} - T & = & 0
\label{gapclose}
\end{eqnarray}
The energy values mentioned above are for $n=1$, which is the bare, unrenormalized scale. The RSRG recursion relations for evaluating the on-site potentials and the hopping integrals are already given in Eqs.~\eqref{recur}. 

Exploiting Eqs.~\eqref{recur} one can obtain the required energy eigenvalues. However, it should be appreciated that with increasing values of $n$ simultaneously satisfying the pair of Eqs.~\eqref{gapclose} will become more and more difficult, and this might lead to a complicated correlation between the values of the system parameters.

\section{Topological issues}
\subsection{The Zak phase} 

We set the lattice constant $a=1$, and $\epsilon=0$ throughout the structure. The Hamiltonian $\hat{\mathcal{H}}(k)$, by construction, exhibits time-reversal symmetry, viz, $\hat{\mathcal{H}}(-k)^\ast = \hat{\mathcal{H}}(k)$. 
For the TVF cluster there is no chiral symmetry, though for the EVF cluster, there is chiral symmetry, observed through the operator, 
\begin{equation}
\hat{\Gamma} = \left[ \begin{array}{cccccccccccccccc}
1 & 0 & 0 & 0 & 0 & 0 \\
0 & -1 & 0 & 0 & 0 & 0\\
0 & 0 & 1 & 0 & 0 & 0\\
0 & 0 & 0 & 1 & 0 & 0\\
0 & 0 & 0 & 0 & 1 & 0 \\
0 & 0 & 0 & 0 & 0 & -1
\end{array}
\right ] 
\label{chiral}
\end{equation}
It can be checked that, $\hat{\Gamma}^{-1}~{\mathcal{\hat H}_{EVF}}(k)~\hat{\Gamma} = - {\mathcal{\hat H}_{EVF}}(k)$ for the case when $\epsilon_a=\epsilon_b=\epsilon_c=0$, the model that we consider here.

Let us discuss the specific cases of an array of $G_1$ clusters of a TVF and an EVF with reference to the effective SSH chain obtained out of a decimation, as shown in Fig.~\ref{RG2SSH}. The conditions for opening (closing) of an energy gap and the required correlation between $T$ and $t$ remain the same, as discussed in the previous subsection.

\begin{figure}[ht]
\centering
(a)\includegraphics[width=0.44\columnwidth]{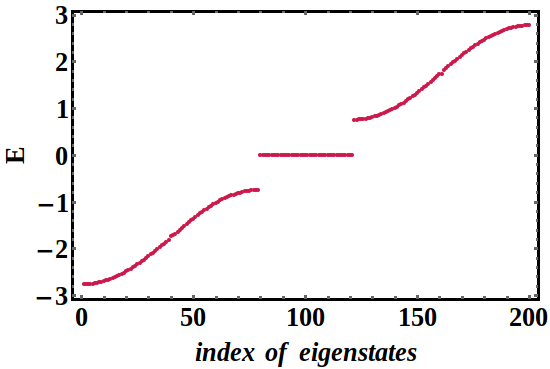}
(b)\includegraphics[width=0.44\columnwidth]{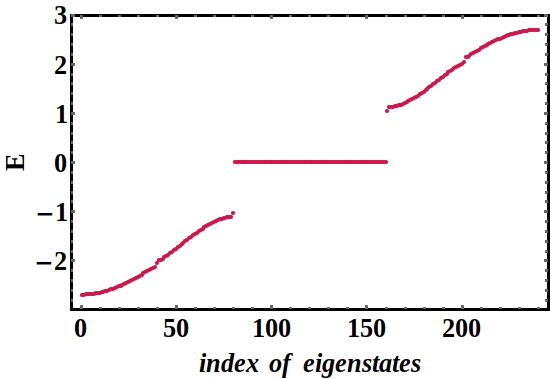}\\
(c)~~\includegraphics[width=0.44\columnwidth]{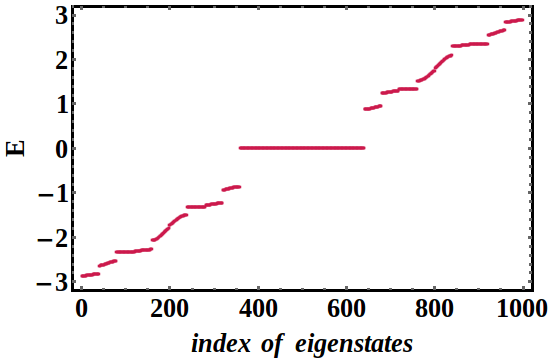}
(d)~~\includegraphics[width=0.44\columnwidth]{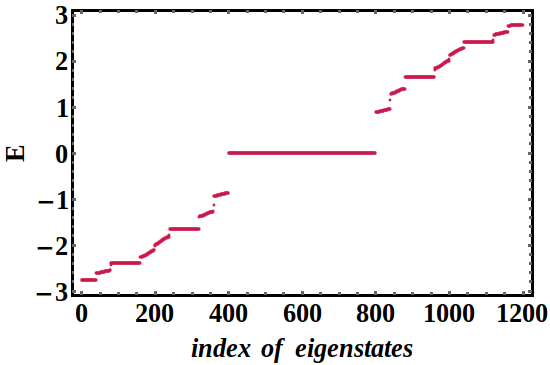}
\caption{(Color online) 
 Distribution of energy spectra against the site index for (a,c) the  traditional Vicsek array  and (b,d) extended Vicsek array  with  $G_1$  and $G_2$ types unit cells  with open boundary condition for $N_x=40$, where $N_x$ denote the number of unit cells taken along the $x$-direction. The values of the parameters are chosen as $t = 1$, $\epsilon_b =\epsilon_c=0$, $T = 1.8$ for (a) and (c), and $t = 1$, $\epsilon_b =\epsilon_c=\epsilon_a=0$, $T = 1.5$ for (b) and (d).}
\label{edgestates}
\end{figure}

The opening of an energy gap that was closed for $v=w$ at the Brillouin zone boundary is only indicative (but not conclusive) of a topological phase transition. We expect a topological invariant to be associated with this phenomenon, a quantity such as the Zak phase~\cite{zak} that flips its quantized value from {\it unity} (in unit of $\pi$) to {\it zero} corresponding to the non-trivial and the trivial insulating phases respectively. Recent experiments have suggested mechanisms for a possible measurement of this topological invariant~\cite{atala}.

The Zak phase for the $n$-th bulk bands is defined as,
\begin{equation}
    Z = -i \oint_{BZ}  \mathcal{A}_{nk}(k) dk
    \label{zak} 
\end{equation}
where $\mathcal{A}_{nk}$ is called the Berry  curvature of the $n$-th Bloch eigenstate, which is again defined as~\cite{asboth},
\begin{equation}
    \mathcal{A}_{n{k}}(k)= \bra{\psi_{n{k}}}\ket{\frac{d\psi_{nk}}{dk}}
\end{equation}

\begin{figure}[ht]
(a)\includegraphics[width=0.44\columnwidth]{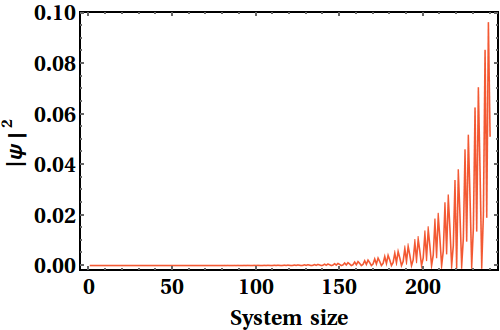}
(b)\includegraphics[width=0.44\columnwidth]{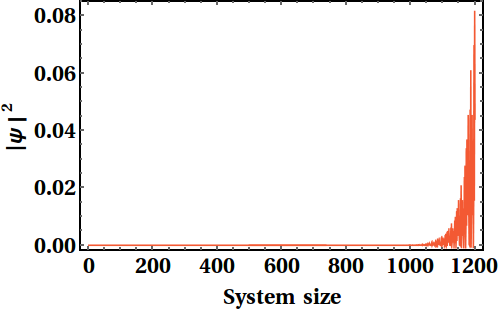}

\caption
 {(Color online) Amplitude distributions of the edge states of EVF with energies (a) $E = \pm 2.04 $ for  $G_1$  type and  (b) $E = \pm 2.03 $  for $G_2$  type unit cell respectively with open boundary condition  $ N_x = 40 $ unit cells. The values of the parameters are chosen as  $t = 1$, $\epsilon_b =\epsilon_c=\epsilon_a=0$, $T = 1.5 $.}
 \label{amplitudeEVF}
\end{figure}

The integral is done along a closed loop in the Brillouin zone.  $\ket{\psi_{nk}}$ is the $n$-th Bloch state. We make use of the Wilson loop approach~\cite{fukui}, that is a gauge invariant formalism. It protects the numerical value of the Zak phase against any arbitrary phase change of Bloch wavefunction.
\begin{figure}[ht]
(a)\includegraphics[width=\columnwidth]{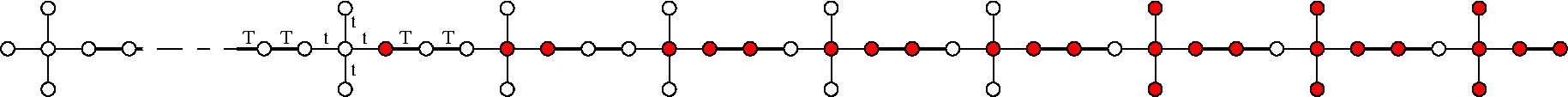}\\\vspace{1cm}
(b)\includegraphics[width=\columnwidth]{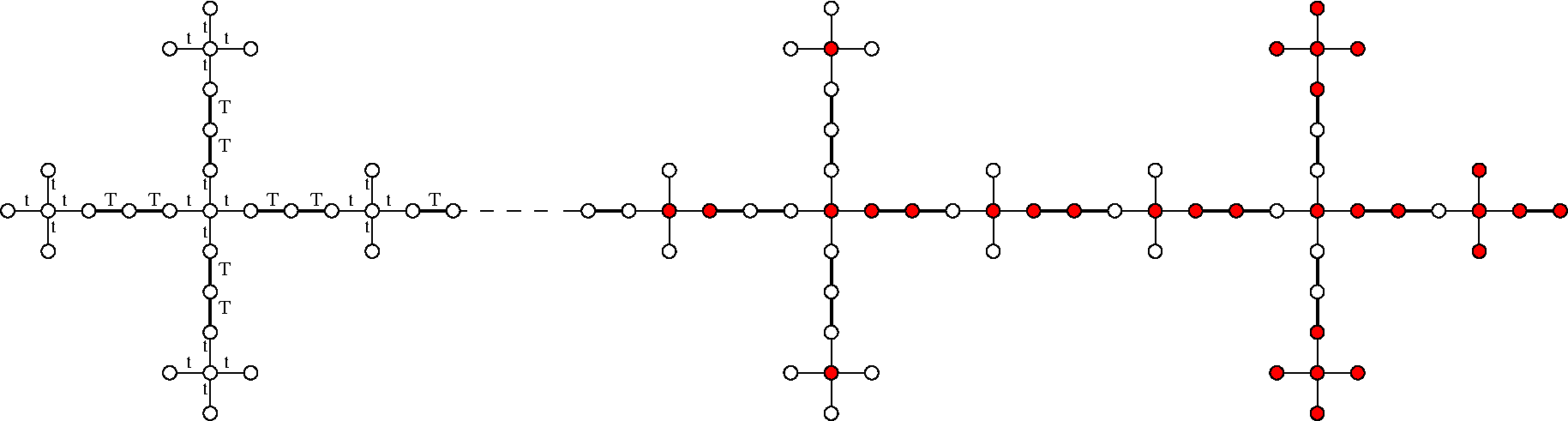}
\caption {(Color online) The magnitudes of the wave functions for the edge states of EVF with energies (a) $E = \pm 2.04 $ for $G_1$ type and (b) $E =\pm 2.03 $ for $G_2$  type unit cell respectively with open boundary condition  $ N_x = 40 $ unit cells. The values of the parameters are chosen as   $t = 1$, $\epsilon_b =\epsilon_c=\epsilon_a=0$, $T = 1.5 $ . The red and white color sites indicate nonzero and almost zero magnitudes of the wave function.}
 \label{amplirealspace}
\end{figure}

In Figs.~\ref{disp1}(a) and (c), corresponding to the TVF cluster ($G_1$) array, we find $Z=0$ for all the bands, including the flat band at $E=0$. The results for the EVF cluster ($G_1$) array in Figs.~\ref{disp1}(b) and (d) are the same, though the sum of the Zak phases of all the bands, including the doubly degenerate flat band at $E=\epsilon_b$ turns out to be an integer multiple of $2\pi$. 

Incidentally, experimental realization of topological FB's in a frustrated kagom\'{e} metal has recently been reported~\cite{kang}. Analyses of flat bands as well as weakly dispersive bands with non-trivial topology have very recently been reported in the literature~\cite{pathak,nandini,bolens}. The present geometry, simpler in nature may be inspiring to experimentalists, to test the existence of topological flat bands in a rectangular mesh.

At this point it is pertinent to raise the question as to whether, for a branching tree-like structure, where the definitions of the {\it bulk} and the {\it boundary} are not clear, especially as the fragmentation increases indefinitely, leading to the true fractal limit, should we expect the bulk-boundary correspondence (BBC) to hold good? The same question was asked previously while investigating the topological states on a Sierpinski gasket fractal~\cite{pai}. However, in the specific cases of periodic arrays of the $G_1$ clusters, the fragmentation of the branches is too small, and the question of a `non-applicability' of the BBC is not meaningful. Rather, the non-existence of a quantized Zak phase in the cases reported above may more appropriately be attributed to the fact that in neither of the cases the axis of the inversion symmetry passes through the center of the unit cells. 

To see if we really can have protected edge states in such systems, we explicitly work out the edges states, both for the TVF and the EVF geometries, using the $G_1$ and $G_2$ clusters and test the robustness of the states against a disorder that eventually mimics a `distortion' in the lattice arrangement. The results are discussed in the following section.

\section{The edge states}

In Fig.~\ref{emergence} we plot the energy spectra as a function of the inter-cluster coupling $T$ for arrays of TVF and EVF, both for the cell sizes $G_1$ and $G_2$. The TVF array does not show any signature of edge states decoupled from the bulk bands. However, the EVF cluster arrays exhibit clear existence of edge states that start getting decoupled from the bulk bands as $T$ exceeds a threshold value. 
As discussed before, the band gaps close for an EVF array at $E=2$ when the inter-cluster hopping is set at $T=\sqrt{2}$. Even a minute deviation from this special value of $T$ opens up a gap around $E=2$, and a `gap-state' appears. This is apparent from Fig.~\ref{emergence} (b) and (d) where we have set $T=1.5$. The gap state decouples from the main bands, and appear at $E\simeq 2$, which is at per with our apprehension. This observation prompts us to examine whether this state resides on the edges of the array as well as the robustness of these modes against a possible disorder in the intra-cluster hopping $t$. This is precisely what we aim at.

First, we plot the amplitudes of the gap states against the lattice site indices. This is shown for arrays of $40$ clusters of EVF with $G_1$ and $G_2$ types of unit cells in Fig.~\ref{amplitudeEVF}(a) and (b). The gap-closing happens at $E=2$ for an EVF array as we set $T=\sqrt{2}$. The energy is now seen to be in the neighborhood of $E=2$, viz, at $E=\pm 2.04$ and $E=\pm 2.03$ for the two cases. The offset in the energy eigenvalue (from $E=2$) is due to a finite de-tuning (to open up a gap) of the value of $T$ from its band-closing value of $T=\sqrt{2}$. The amplitudes of the wave function in both the cases reside towards one end of the chain, decaying exponentially in the bulk.

To see the distribution of amplitudes in real space we also plot $|\psi|^2$ throughout the lattice. This is depicted in Fig.~\ref{amplirealspace} (a) and (b). For an array of $G_1$ cells (in (a)), the gap-state at $E=2.04$ for the $G_1$ array have their amplitudes spread over the final seven unit cells towards the right side of the chain. Interestingly, for the $G_2$ array, as seen in Fig.~\ref{amplirealspace}(b), the unit cell expands in the transverse direction, and the amplitudes get pinned over the branches along the $y$-axis. The number of unit cells hosting the gap-states are now much less, and we anticipate that, with $G_3$ and $G_4$ cells forming the array, the amplitudes will be localized over still smaller clusters at one end of the array. These are edge states.
\begin{table}[ht]
\centering
\caption{Robustness of the edge states against hopping disorder for an EVF array}
\vspace{0.5cm}
\begin{tabular}{|p{1.5cm}|p{2cm}|p{2cm}|p{1.5cm}|p{1cm}|}
 \hline
 \multicolumn{5}{|c|}{\textbf{ Extended Vicsek : Unit Cell $G_1$ } }\\
 \hline
 Whether disorder applied & $N_{cell}^{dis}$ / $N_{cell}^{tot}$  & $N_{bond}$ &  $t+\delta$ & $E_{edge}$ \\
 \hline
 No     &  0/40   & 0 &  0 & ±2.04\\
 \hline
 Yes   &   40/40  & 1 & 1 to 1.02 & ±2.04 \\
 \hline
 Yes   &   40/40  & 2 & 1 to 1.01 & ±2.04\\
 \hline
 Yes   & 40/40  &  3 & 1 to 1.01 & ±2.04\\
 \hline
 Yes   & 40/40 & 4 & 1 to 1.01 & ±2.04\\
 \hline
 \multicolumn{5}{|c|}{ \textbf{Extended Vicsek : Unit Cell $G_2$}} \\
 \hline
 No  &  0/40  &  0 &  0 & ±2.03\\
 \hline
 Yes   &   40/40  & 4 & 1 to 1.1 & ±2.03\\
 \hline
 Yes  & 40/40 & 5 & 1 to 1.1 & ±2.03\\
 \hline
 Yes   & 40/40 & 8 & 1 to 1.05 & ±2.03\\
 \hline
 Yes   & 40/40 & 10 & 1 to 1.03 & ±2.03\\
 \hline
  yes   &   40/40  & 12 & 1 to 1.02 & ±2.03\\
 \hline
 Yes   & 40/40 & 14 & 1 to 1.02 & ±2.03\\
 \hline
 Yes   & 40/40 & 16 & 1 to 1.01 & ±2.03\\
 \hline
 Yes   & 40/40 & 20 & 1 to 1.01 & ±2.03\\
 \hline
\end{tabular}
\label{stability}
\end{table}

We now test the robustness of the edge states in an EVF array against disorder. In Table~\ref{stability} we summarize the results of applying disorder to a $40$-cell periodic array of $G_1$ clusters of an EVF first, and then $G_2$. The disorder is implemented by replacing the intra-cell hopping integral $t$ by $t+\delta$, and then choosing $\delta$ randomly from a window. The choice is specified in the fourth column in Table~\ref{stability}. The disorder mimics `deformation' in the structure. Any choice of $T \ne \sqrt{2}$ would open up a gap in the spectrum around $E=2$. We have chosen $T=1.5$ in units of $t$, and observe a clear emergence of a gap-state around $E=2 \pm \delta$, where $\delta \sim 0.03$, caused by the detuning of $T$. 

A close look at the Table~\ref{stability} unveils several interesting features shown by the states reported. For an EVF array with a $G_1$ unit cell, the energy of the gap state remains unchanged even when we introduce disorder in every unit cell comprising the array. This is shown in the table as the ratio $N_{cell}^{dis}/N_{cell}^{tot}$. The number of bonds in a cell, where disorder is introduced, is denoted as $N_{bond}$. The table shows that the state at $E=2.04$ is robust in energy as well as its distribution, when the intra-cell hopping in each cell deviates (though marginally) from its initial value of $t=1$. The edge state in this case can be said to be {\it weakly protected} against disorder.

The situation looks better when the size of the unit cell increases to $G_2$. In this case, every unit cell has 30 bonds, and we have given disorder up to to 20 (twenty) bonds in a cell to see if the edge state deviates in energy or distribution. We see that tampering the hopping for up to five (5) bonds in a cell, going from $t=1$ to $t=1.1$ does not affect the stability of the gap state at all. With increasing number of bonds per unit cell where disorder is introduced, the width of disorder of course, is a bit smaller. But, the disorder is spread all over the bulk of the system.

From the above observations we conclude that, edge states indeed appear for an EVF array, and these are robust, at least in a weak-disorder sense, against bulk disorder injected into the lattice. 

\section{Concluding Remarks}
In conclusion, we have studied arrays of branched structures derived out of the basic cells of a traditional Vicsek fractal, and an extended Vicsek fractal. We have given an exact prescription to obtain a multitude of flat bands when the unit cells of each of these fractal arrays grow in size, tending to a full Vicsek geometry in the thermodynamic limit. We have also investigated the topological properties of such arrays, in particular, the topological invariant (the Zak phase) and the edge states. There are edge states, and these edge states for an extended Vicsek fractal array turn out to be robust against disorder, even when the disorder is spread throughout the bulk. The disorder in general, needs to be weak, especially when more and more bonds are detuned from their initial values. The apparent anomalous observation of protection and absence of quantized Zak phase values may be attributed to the fact that the axis of inversion symmetry is displaced from the centers of the unit cell in such structures.

Finally, the method can be continued to extract information regarding the flat bands, the edges and all that, for arbitrarily large unit cell structure, simply by replacing the basic on-site potentials and the hopping integrals by their renormalized values. The number of renormalization steps $n$ will indicate how big a unit cell we started out with, and in principle, it will be possible to have an understanding of the topological properties of a Vicsek fractal as $n \rightarrow \infty$.

\section{Acknowledgments}
A. M. worked on this problem during her formal stay at the University of Kalyani, and acknowledges DST for providing her with an INSPIRE Fellowship $[IF160437]$. S. B. is thankful to Government of West Bengal for the SVMCM Scholarship. A. M. thanks Presidency University for providing the computational facility.

\section{Appendix}

The recursion relations for a EVF are given below. For simplification, we choose to write $ (E-\epsilon_{b,n}) = E_{b,n}$, $(E-\epsilon_{c,n}) = E_{c,n}$ and $(E-\epsilon_{a,n}) = E_{a,n}$.

Decimation of a selected set of sites, as shaded out in Fig.~\ref{vicsek}(c), leads to the set of recursion relations,

\begin{eqnarray}
\epsilon_{b,n+1} & = & \epsilon_{b,n} + \frac{t_{n}^2}{E_{c,n}-\frac{2 t_{n}^2}{E_{b,n}}-\frac{t_{n}^2}{E_{b,n}-\frac{T_{n}^2}{E_{a,n}-\frac{T_{n}^2}{E_{b,n}}}}}
\nonumber \\
\epsilon_{c,n+1} & = & \epsilon_{c,n}+\frac{4 t_{n}^2} {E_{b,n}-\frac{T_{n}^2}{E_{a,n}-\frac{T_{n}^2}{E_{b,n}-\frac{t_{n}^2}{E_{c,n}-\frac{2 t_{n}^2}{E_{b,n}}}}}}
\nonumber \\
\epsilon_{a,n+1} & = & \epsilon_{a,n} \nonumber \\
t_{n+1} & = & \frac{t_{n}^3 T_{n}^2} {(E_{c,n}-\frac{2 t_{n}^2}{E_{b,n}})(E_{a,n} E_{b,n}^2 - 2 T_{n}^2 E_{b,n}) - t_{n}^2(E_{a,n} E_{b,n} - T_{n}^2)} 
\nonumber \\
T_{n+1} & = & T_{n}
\label{recurevf}
\end{eqnarray}
for RSRG steps $n \ge 0$. Here, it is implied that, at the iteration step $n=0$, all the parameters assume their initial values, as discussed in the text.

\end{document}